\documentclass[runningheads]{svmult}

\usepackage{makeidx}   
\usepackage{graphicx}  
\usepackage{subeqnar}  
\usepackage{multicol}  
\usepackage{physprbb}  
\makeindex             

\begin{document}

\def\MSUN{\rm M_{\odot}}
\def\RSUN{\rm R_{\odot}}
\def\MSUNYR{\rm M_{\odot}\,yr^{-1}}
\def\MDOT{\dot{M}}

\newbox\grsign \setbox\grsign=\hbox{$>$} \newdimen\grdimen
\grdimen=\ht\grsign
\newbox\simlessbox \newbox\simgreatbox
\setbox\simgreatbox=\hbox{\raise.5ex\hbox{$>$}\llap
     {\lower.5ex\hbox{$\sim$}}}\ht1=\grdimen\dp1=0pt
\setbox\simlessbox=\hbox{\raise.5ex\hbox{$<$}\llap
     {\lower.5ex\hbox{$\sim$}}}\ht2=\grdimen\dp2=0pt
\def\simgreat{\mathrel{\copy\simgreatbox}}
\def\simless{\mathrel{\copy\simlessbox}}

\title*{Time Variability of Low Angular Momentum Flows Accreting onto Black
Holes: A Natural Mechanism For Radiation Flaring }
\toctitle{Time Variability of Low Angular Momentum
\protect\newline Flows Accreting onto Black Holes:
\protect\newline A Natural Mechanism For Radiation Flaring}
%
%
\titlerunning{Accretion of Low Angular Momentum Material onto Black Holes}
%
\author{Daniel Proga\inst{1, }\inst{2}}
\authorrunning{Daniel Proga}
%
%
\institute{JILA, University of Colorado, Boulder CO 80309, USA
\and Present address: Department of Astrophysical Sciences, 
Princeton University, 
Peyton Hall,
Princeton NJ 08544, USA}

\maketitle              

\begin{abstract}
We  present results from our magnetohydrodynamical
simulations of accretion flows onto black holes. Our main focus is 
the interplay between inflows and related outflows. We 
consider applications of such flows to the Galatic center and 
low luminosity active galactic nuclei.
\end{abstract}

\section{Introduction}
Dynamical evidence suggests that the nonluminous matter within
0.015 pc of the Galactic center has a mass of
$\approx 2.6\times 10^{6}\ M_\odot$ (e.g.,~\cite{Ge},~\cite{Gh}).
This matter is associated 
with Sgr~A$^\ast$, a bright, compact radio source~\cite{BB},
and provides very compelling evidence
for the existence of a supermassive black hole (SMBH).
Observations of Sgr~A$^\ast$ in X-ray and radio bands reveal a luminosity 
substantially below the Eddington limit, 
$L_{\rm Edd}=3 \times 10^{44}$ erg s${}^{-1}$.
For example, {\it Chandra} observations show  
a luminosity in 2-10 keV X-rays of $\approx 2\times
10^{33}$ erg s${}^{-1}$ (ten orders of magnitude 
below $L_{\rm Edd}$)~\cite{Ba01}.
{\it Chandra} observations also revealed
an X-ray flare rapidly rising to a
level about 45 times as large, lasting for only $\sim 10^4$~s,
indicating that the flare must originate near the black hole~\cite{Ba01}.

It is thought that the Sgr~A$^\ast$ radiation is due to gas accretion
onto the SMBH. Estimates for the accretion luminosity, $L$, rely on
assumptions about the mass accretion rate, $\MDOT_a$, and the efficiency
of transforming the gas energy into radiation, $\eta$ (i.e.,
$L= \eta c^2 \MDOT_a$).
Both $\MDOT_a$ and $\eta$ are uncertain and there is no
generally accepted model which could explain the low luminosity of
Sgr~A$^\ast$ by predicting low enough $\MDOT_a$ or $\eta$, or both.
For example,  Coker \& Melia~\cite{CM} 
 estimated a rate of $10^{-4}~\MSUNYR$
from Bondi-Hoyle accretion of winds from nearby stars, while 
Quataert, Narayan, \& Reid ~\cite{QNR} estimated the Bondi capture rate 
of $3 \times 10^{-5}~\MSUNYR$. On the other hand,  
the best fit spectral models of ~\cite{MLC} and ~\cite{YQN} 
have $\MDOT_a= 10^{-10}~\MSUNYR$ and $10^{-8}~\MSUNYR$, respectively.    
Quataert, Narayan, \& Reid~\cite{QNR} argue that the low
luminosity requires $\MDOT_a$ to 
substantially sub-Eddington and sub-Bondi at large radius.
However, the rate at which  mass is captured into
the accretion flow at large radii, the mass supply rate, does not
necessarily have to be
the  same as the rate at which mass is accreted onto a black hole.

\begin{figure}[bht]
\begin{center}
\includegraphics[width=.5\textwidth, angle=90 ]{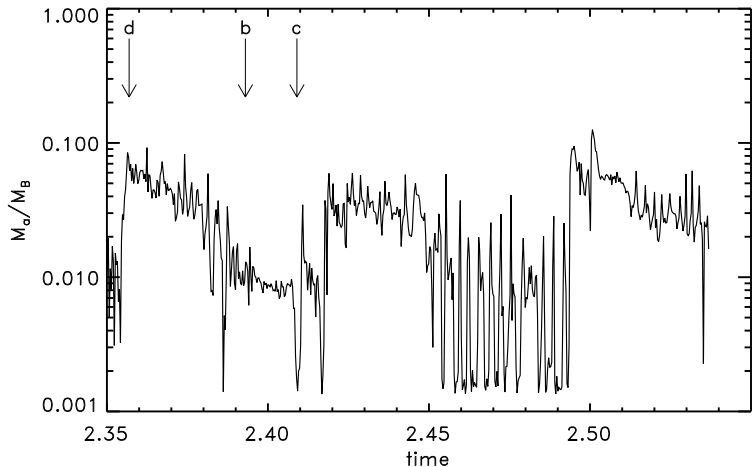}
\end{center}
\caption[]{
Late time evolution of the mass accretion rate in units of the
Bondi rate, for the fiducial model of MHD accretion flow
in ~\cite{PBb} (their model D).
Time is in units of the Keplerian orbital time at the Bondi radius,
which for this simulation was set at 1000 times the black hole
radius.
The figure shows in detail $\MDOT_a$ as a function of time
toward the end of the simulation.
Vertical arrows mark times corresponding to three (out of four) generic
states
of accretion:
accretion is dominated by low-$l$ material which managed to reach
the inner boundary despite a blocking  corona and outflow from the torus
(arrow~d), accretion occurs only through the torus (arrow~b) and
no  torus accretion but only very weak accretion through
a very low density magnetized polar cylinder (arrow c). See Figs.~2 and 3 
for the density maps and the velocity fields of the inner flow corresponding
to the times marked.}
\label{eps1}
\end{figure}

\begin{figure}[!htb]
\begin{center}
\includegraphics[width=0.80\textwidth]{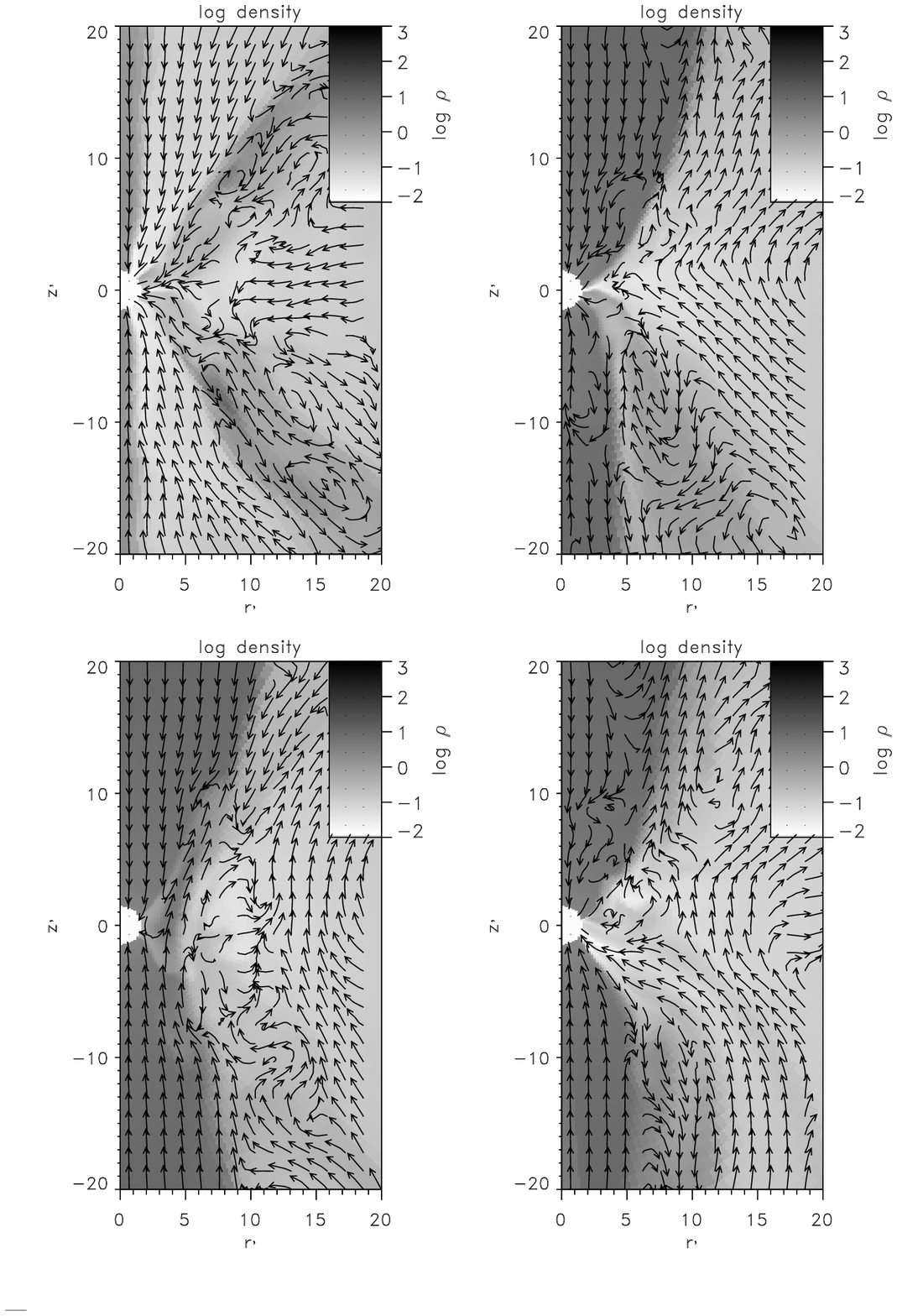}
\caption{Maps of logarithmic density overplotted by the direction
of the poloidal velocity. This figure compares the inner flow
in four different accretion states in our MHD simulations
(Fig.~8 in ~\cite{PBb}).
We express the length scale in units of the black hole radius.
The top left panel presents the two-dimensional structure
near the beginning of simulations at $t=0.22$ (not shown in Fig.~1).
At this time, accretion onto the black hole occurs through
both the torus and the polar funnel.
The top right panel  presents an example of an inner flow where accretion
occurs only through the torus at $t=2.39$ (marked by arrow~b in Fig.~1).
The bottom left  panel  presents an example of an inner flow where
there is no  torus accretion but only very weak accretion through
a very low density magnetized polar cylinder at $t=2.41$
(marked by arrow~c in Fig.~1, note that this  state  is very short-lived). 
Finally, the bottom right  panel  presents an example of an inner flow where
accretion is dominated by low-$l$ material (i.e., the inward stream of gas
directed present in the lower half of the panel).  This material 
managed to reach the inner boundary despite a blocking  
corona and outflow from the torus at $t=2.36$ (marked by arrow~d in Fig. 1).
}
\end{center}
\end{figure}

Many aspects of spectral models for Sgr~A$^\ast$ follow from the simple
scaling laws of black hole accretion and will be present in any model,
regardless of the detailed dynamics. For a steady state model,
an impressive range of spectra can be generated with the adjustment of several
free parameters: the ratio of electron to ion temperature, the
magnetic pressure, the radial density profile, the mass-accretion
rate, and the mass-loss rate (e.g., ~\cite{QN}).
Thus it is difficult to tightly constrain parameters of a steady state  model.
However, our recent magnetohydrodynamical (MHD)
simulations present a way of constraining the
accretion flow models by studying time variability~\cite{PBb}.
As we will describe in next section, the simulations show that
the inner accretion flow quasi-periodically changes both quantitatively
(e.g., $\MDOT_a$ changes by $\simgreat 1$ order of magnitude) and
qualitatively (e.g., the inner flow can be dominated by an equatorial
accretion torus or a polar inflow). 

\section{Results From MHD Simulations Of Accretion Flows}

We have performed axisymmetric, MHD, time-dependent simulations of slightly 
rotating accretion flows onto black holes ~\cite{PBb}. Our simulations are
complementary to previous MHD simulations which considered strongly
rotating accretion flows and started from a rotationally supported torus  
(e.g.,~\cite{Ba03},~\cite{Di}, ~\cite{Fa}).
We attempt to mimic the boundary conditions of classic Bondi accretion
flows as modified by the introduction of a small, latitude-dependent
angular momentum at the outer boundary, a pseudo-Newtonian gravitational 
potential and weak poloidal magnetic fields. A weak radial magnetic field 
and distribution of the specific angular momentum, $l$, with latitude allow 
the density distribution at infinity to approach spherical symmetry. 
Therefore our outer boundary conditions are consistent with
X-ray images taken with {\it Chandra} which show that the gas distribution in 
the vicinity of SMBHs is close to spherical. 

We find that the material with high $l$ forms an equatorial torus which can
accrete onto the black hole because of magnetorotational instability (MRI).
The magnetized torus produces a corona and an outflow.
The latter two can be strong enough to prevent
accretion of low-$l$ material through the polar regions
(the source of accretion in the hydrodynamical inviscid case ~\cite{PBa}).
We find that the net mass accretion rate through the torus is lower 
than the Bondi rate and also lower than $\MDOT_a$ in the HD inviscid case.

The accreting torus is the crucial component of our accretion flow.
After an initial transient, the inner flow usually consists of a turbulent, 
gas pressure-dominated MHD torus with an outflow or corona or both, 
bounded by a magnetic pressure-dominated polar flow.
The accretion through the torus is highly variable. In fact, it  
could be stopped for a relatively short time, by the accumulated poloidal 
magnetic flux that builds up during accretion and truncates the torus. 
However, we observe that even  when the torus is truncated, there is inflow 
of material inside the torus and its mass and pressure build up.
Consequently, the magnetic field is quickly pushed
inward by the torus and  the gas from the torus can again fall
onto the black hole (note short-lived `dips' and 'spikes' in the time
evolution of $\MDOT_a$ shown in Fig.~1). Rapid time variability
appears to be typical for  MHD turbulent torii. 
Generally, we find that the properties of the inner flow (e.g., the radial
density profile in the torus, properties of the corona and outflow)
are very similar to those presented by ~\cite{SP} and ~\cite{HB}
despite  using different initial and outer boundary conditions. 

What is new in our simulations is the fact that the torus accretion
can be supplemented or even replaced by  stream-like accretion
of the low-$l$ material occurring outside the torus (e.g., Figs. 1, 2 and 3).
When this happens, $\MDOT_a$ sharply increases and gradually decreases 
in a quasi-periodic manner. $\MDOT_a$ due
to this 'off torus' accretion can be one order of magnitude larger than
that due to the torus (Fig.~1). The off torus accretion is a consequence
of our outer boundary and initial conditions which introduce
the low-$l$ material to the system. This material can reach the black hole
because the torus corona and outflow are not always
strong enough to push it away.

One can expect that in the vicinity of SMBHs some gas has very little
angular momentum and could be directly accreted. Such a situation
occurs likely at the Galactic center
where a cluster of young, massive stars losing
mass surrounds a SMBH. 
We propose then that the variability observed in Sgr~A$^\ast$
(i.e., flares) could be due to the interplay between
gas with a range of specific angular momentum similar to that found
in our simulations.

\begin{figure}[!htb]
\begin{center}
\includegraphics[width=0.800\textwidth]{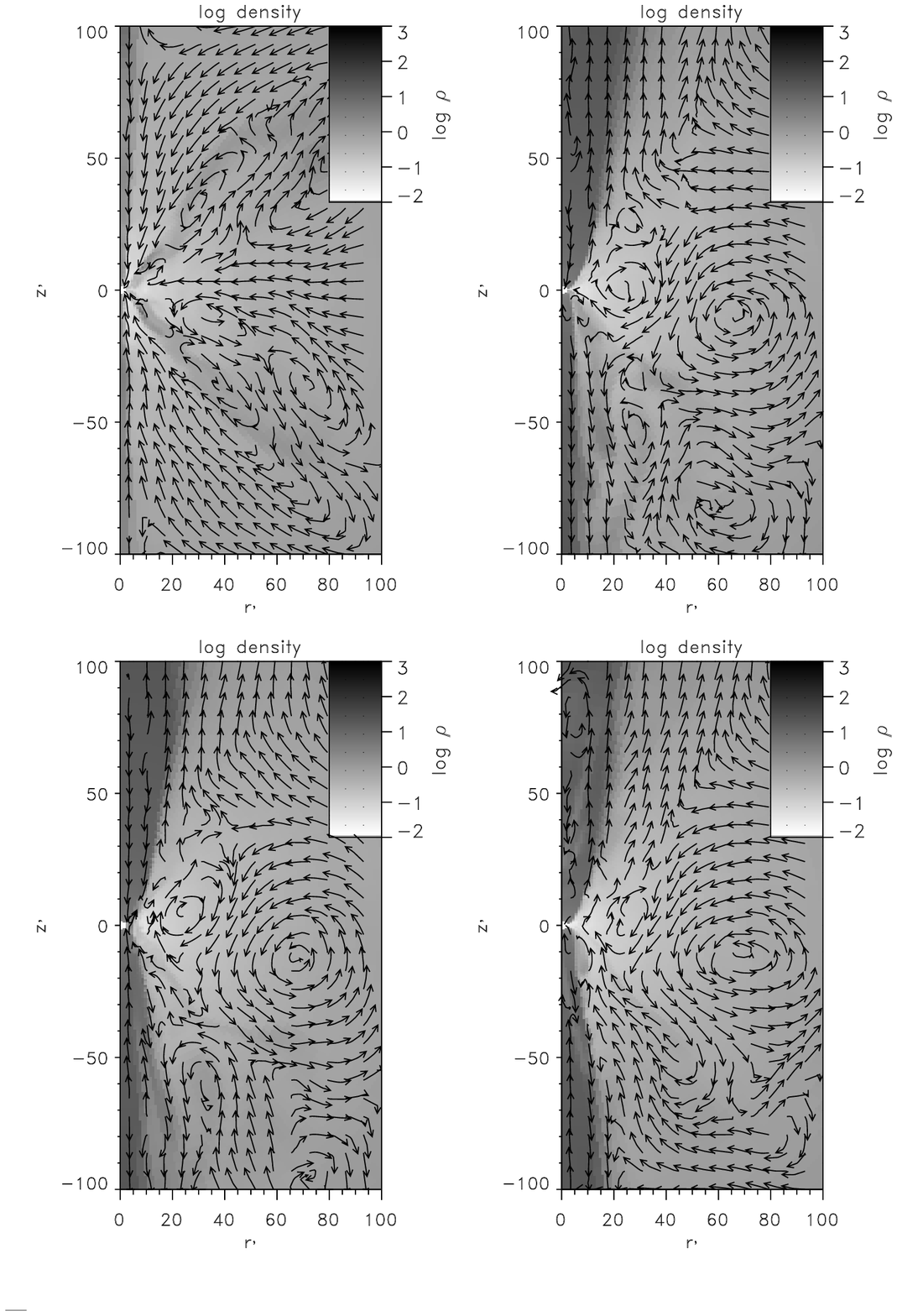}
\caption{As Fig. 2 but with the $r'$ and $z'$ ranges 
increased by a factor of 5. Note the large scale, polar 
double-sided outflow from the
thick equatorial torus in the top right and bottom left panels. 
The bottom right panel shows an accretion state where  the outflow below
the equator is
replaced by an inflow. We propose that switching between this state 
and that presented in the top right panel (i.e., in Fig.~1, 
the states marked by arrows d and b, respectively) can explain radiation 
flaring in Sgr~A$^\ast$. 
}
\end{center}
\end{figure}

\section{Conclusions}

Our main results can be summarized as follows:

$\bullet$ The properties of the accretion flow depend on an equatorial torus.

$\bullet$ Accretion can be via the torus due to MRI and 
via the polar funnel where material has zero or low angular momentum.

$\bullet$ The torus outflow and corona are natural mechanisms to narrow 
or even totally close the polar funnel for the accretion of  low-$l$ material.

$\bullet$ The net accretion rate  for the MHD, is lower than for 
the Bondi flow and the HD inviscid flow. 

$\bullet$ Time variability of the inner flow may explain light curves 
in Sgr~A$^\ast$. In particular, we propose that the X-ray flares can 
be explained by quasi-periodic, off-torus accretion of  the low $l$ material.

We acknowledges support from
NASA LTSA grant NAG5-11736.
We also acknowledge support provided by NASA through grant
HST-AR-09947 from the Space Telescope Science Institute, which is operated
by the Association of Universities for Research in Astronomy, Inc.,
under NASA contract NAS5-26555.

%

\end{document}